# Highly reproducible low temperature scanning tunnelling microscopy and spectroscopy with *in situ* prepared tips

Andres Castellanos-Gomez[1,2,*] Gabino Rubio-Bollinger[1,3,*], Manuela Garnica[1,4], Sara Barja[1,4], Amadeo L. Vázquez de Parga[1,3,4], Rodolfo Miranda[1,3,4] and Nicolás Agraït [1,3,4].

[1] Departamento de Física de la Materia Condensada (C–III). Universidad Autónoma de Madrid, Campus de Cantoblanco, 28049 Madrid, Spain.
[2] Kavli Institute of Nanoscience, Delft University of Technology, Lorentzweg 1, 2628 CJ Delft, The Netherlands.
[3] Instituto Universitario de Ciencia de Materiales "Nicolás Cabrera". Campus de Cantoblanco, 28049 Madrid, Spain.
[4] Instituto Madrileño de Estudios Avanzados en Nanociencia IMDEA-Nanociencia, 28049 Madrid, Spain.

E-mail: a.castellanosgomez@tudelft.nl, gabino.rubio@uam.es .

An *in situ* tip preparation procedure compatible with ultra-low temperature and high magnetic field scanning tunneling microscopes is presented. This procedure does not require additional preparation techniques such as thermal annealing or ion milling. It relies on the local electric-field-induced deposition of material from the tip onto the studied surface. Subsequently, repeated indentations are performed onto the sputtered cluster to mechanically anneal the tip apex and thus to ensure the stability of the tip. The efficiency of this method is confirmed by comparing the topography and spectroscopy data acquired with either unprepared or *in situ* prepared tips on epitaxial graphene grown on Ru (0001). We demonstrate that the use of *in situ* prepared tips increases the stability of the scanning tunneling images and the reproducibility of the spectroscopic measurements.

## 1. Introduction

Since its invention, the scanning tunneling microscope (STM) has been employed as a very powerful tool to characterize the topography of surfaces with atomic resolution. Additionally, an STM can be used to study the local electronic properties of surfaces by performing scanning tunneling spectroscopy (STS). The measured spectra, however, depend also on the exact atomic arrangement of the atoms involved in the tunneling process at the tip apex. Therefore, the stability and chemical composition of the tip apex have a strong impact on both the resolution of the STM images and reproducibility of spectroscopic data.

Many of the environmental conditions required in some STM experiments (such as low temperature, high magnetic field or ultrahigh vacuum (UHV)) may hinder the preparation, handling and substitution of tips. Therefore, reproducible *in situ* tip preparation and characterization procedures become crucial. Nevertheless, most of *in situ* tip preparation procedures require dedicated setups such as ion milling, argon sputtering or UHV annealing [1-3] which are typically not available in some STM setups (especially in those operated at ultralow temperatures and high magnetic fields). Other common *in situ* preparation procedures rely on intentionally crashing the tip against the substrate to coat the tip with the substrate material [4]. However, this procedure has been only applied for a limited number of substrates such as gold, silver, lead, aluminum or copper.

Here we describe an *in situ* tip preparation procedure compatible with STMs operated at cryogenic temperatures and high magnetic fields. The tips are prepared by electric-field-induced deposition of material from the tip on the studied surface. In order to increase the mechanical stability of the tips, repeated indentations onto the sputtered material are performed to mechanically anneal the tip apex. *In situ* prepared tips have been employed to study the topography and the local electronic properties of epitaxial graphene grown on Ru (0001) at low temperature.

## 2. Scanning tunnelling microscopy and spectroscopy with unprepared tips

A graphene sample has been epitaxially grown onto Ru (0001) in UHV [5, 6], then exposed to air and transferred to a $^3$He cryostat equipped with a 9 T superconducting magnet. We use a homebuilt low temperature STM, operating at a base temperature of < 300 mK, similar to the one described in ref.[7]. The low reactivity of the graphene surface [8] makes this transfer possible without a noticeable degradation of the surface quality due to atmospheric exposure. The STM tip has been first prepared *ex situ* by cutting a high purity (99.99 %) gold wire with scissors. Although these mechanically cut tips are rather usual in STM, the STM imaging and STS measurements are quite sensitive to the exact details of the tip apex [9]. Figure 1a shows an STM topography, measured at T < 300 mK, of the rippled structure of graphene epitaxialy grown on Ru (0001), which presents a Moiré pattern, with a periodicity of 3 nm, superimposed to the atomic periodicity. The Moiré pattern is originated by the lattice parameter mismatch between the graphene overlayer and the ruthenium surface [10] and also causes an inhomogeneous graphene-ruthenium interaction modulating the electronic properties of graphene [6, 10]. In Figure 1a the atomic periodicity is only resolved during the first half of the image and, then, there is a sudden decrease of the resolution probably due to a change in the tip (*ex situ* prepared) apex structure. Figure 1b shows three different spectroscopic traces measured on top of the Moiré ripples. These traces are characterized by the lack of reproducibility. Even, some of the spectroscopic traces show evidence of Coulomb blockade behavior.

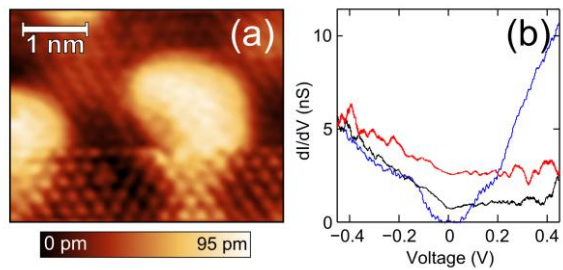

**Figure 1**: (a) STM topography showing a sudden decrease in the resolution due to a structural change in the tip apex ($V_{sample}$ = 1 V; $I$ = 1.5 nA). (b) Spectroscopic traces measured on top of three Moiré ripples with an unprepared STM tip. Notice the lack of reproducibility of the d$I$/d$V$ curves.

## 3. *In situ* tip preparation procedure

To overcome the lack of reproducibility due to the unstable tip apex structure, we have developed an *in-situ* tip preparation procedure to fabricate reproducible STM gold tips. First we use electric-field-induced deposition of tip material to form a gold nanocluster on the graphene surface [11, 12] (see Figure 2a). Large nanoclusters can be reliably deposited by applying a short pulse (0.05 sec.) of + 6V to + 9V to the sample while the tunneling tip is close to the surface (tunneling resistance 100 MΩ) with the feedback turned off. We find that using a pulse of 0.05 sec and + 9 V is optimal to deposit gold clusters large enough to prepare our STM tips with an almost 100% yield. Previous works demonstrate, on the other hand, that different voltage and pulse-time conditions can be employed to control the size of the deposited nanodots even for other tip materials such as aluminum or silver [13, 14]. In order to determine whether the nanodot has enough material to proceed with the tip preparation procedure, we characterize the topography of the cluster by STM imaging to ensure that enough tip material has been transferred (see the 3D topographic reconstruction in the third panel of Figure 2a). The gold nanodots prepared by applying a pulse of 0.05 sec and + 9 V to the tip are typically 20 nm in diameter and 3 nm in height which is enough material for the tip preparation procedure. Interestingly, we found the geometry of the deposited nanodots is rather reproducible even for different initial tips (with different tip apex geometry).

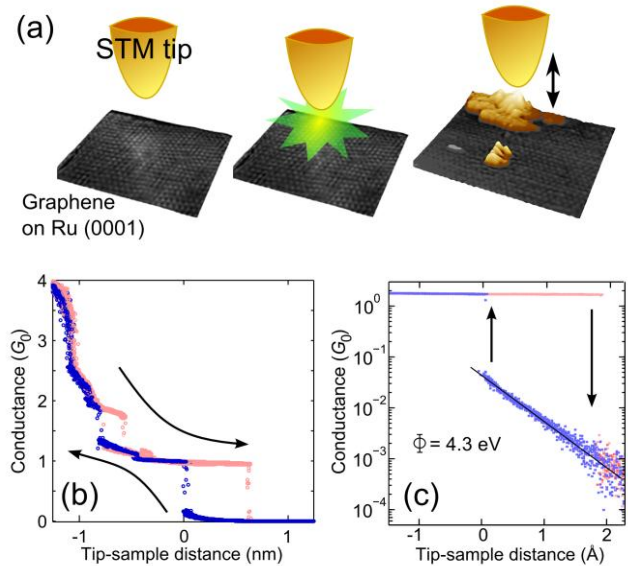

**Figure 2**: (a) Schematic diagram of the *in situ* tip preparation procedure. First, the STM gold tip is brought close the surface until a tunneling current is established. Second, a voltage pulse (+9 V during 0.05 sec) is applied to the sample to deposit a portion of the tip apex on the sample. The last panel in (a) show the 3D reconstruction of the STM topography of a gold cluster sputtered from the tip. Finally, the tip is indented in the gold nanocluster several times to mechanically anneal the tip apex. When the conductance *vs.* distance traces are reproducible a very stable and clean one atom gold contact is formed. (b) 16 Conductance *versus* tip-sample distance traces measured during the nanoindentations onto the gold cluster. (c) Formation and breaking of several hundreds of single-atom contacts between the tip and the cluster. The conductance of the contact is nearly constant at one $G_0$ and when the contact is broken the conductance decays exponentially with the distance. The tunneling barrier height associated to this exponential decay is 4.3 eV. A logarithmic scale has been used to facilitate the determination of the tunneling barrier height..  In this representation the single-atom contact conductance ($G_0$) appears as a featureless horizontal line.

Finally we perform several indentation/retraction cycles with the STM tip on the nanocluster measuring simultaneously the conductance to follow the evolution of the nanocontact [15, 16][17]. The measurement of the conductance allows one to infer the shape and the size of the gold-gold nanocontact and thus to accurately control the indentation/retraction procedure [16, 18]. These cycles can be carried out very fast, measuring thousands of traces in few minutes (time per trace is 0.25 sec), using the standard current *vs.* distance mode available in most of STM microscopes (usually called *IZ* spectroscopy mode). Initially, the tip is indented to get a large nanocontact (with a conductance of about 10 $G_0$, being $G_0 = 2e^2/h$ the conductance quantum) and thousands of indentation/retraction cycles with a tip displacement of about 2 nm are performed. Then the tip is retracted by 0.1 – 0.2 nm and another set of indentation/retraction cycles are carried out. This process is repeated until the nanocontact can be broken and formed during the cycles. A hysteretic behavior is observed (Figure 2b and 2c) when the nanocontact yields abruptly due to sudden changes of atomic configuration, relaxing the elastic stress accumulated during the cycle [19]. The conductance during the last stage of the contact is close to $G_0$ [20] and the tunneling barrier height measured after the gold nanocontact is broken is 4.3 ± 0.5 eV (Figure 2c) indicating that a clean monoatomic sharp gold tip apex has been fabricated. Notice that the tip apex termination plays the most important role in the tunneling process. Therefore fabrication of tips terminated by a single gold atom in a controlled way increases the reproducibility of the scanning probe experiments. It is also important to notice that the presence or absence of conductance plateaus in the indentation/retraction cycles can be used to infer whether the nanocontact is clean or contaminated respectively. In the case of a very contaminated initial tip, it could be necessary to repeat the electric-field-induced deposition step several times to get a clean gold nanocontacts. In our experiments, on the other hand, we always found that one electric-field-induced deposition step was enough to achieve clean gold nanocontacts.

In metallic nanocontacts, dislocations are not energetically favorable and they are thus quickly expelled towards the bulk due to the repeated plastic deformation during the indentation cycles, in a so called mechanical annealing process [21]. Nanocontacts, fabricated by repated indentation, can sustain larger yield stress before relaxation than macroscopic contacts [21] because new dislocations would have to be nucleated for plasticity to proceed. Similar mechanical behavior have been observed in microfabricated metallic pillars in where mobile dislocations escape from the crystal at the nearby free surfaces during the deformation process, leading to a state of dislocation starvation and increasing the strength of the pillars [22, 23].

Here we employed repeated indentation cycles to perform a mechanical annealing process, expelling the defects from the gold nanocontact and strengthening it [16]. This mechanical annealing process forms a highly crystalline [24, 25] and stable nanocontact [25] revealed by the reproducibility of the conductance traces after several hundreds of cycles [26, 27] (Figure 2b). Moreover, a recent work combining electronic transport measurements and molecular dynamics [25] showed that after breaking a mechanically annealed nanocontact a stable and atomically well-defined tip is obtained. The stability of the fabricated tips, however, depends on the thermal diffusion of atoms in the tip apex and thus the maximum stability can be achieved at cryogenic temperatures where the tip apex atoms can be stable for several hours.

The mechanical annealing procedure can be applied to different tip materials [16, 28] (such as Ag, Al, Pb, Sn, Rh, Pd, Ir, Pt) with different properties, which broads the range of experiments that can be carried out with this *in situ* prepared tips.

**4. Scanning tunnelling microscopy and spectroscopy with tips prepared *in situ***

Once the tip has been prepared *in situ*, the position of the tip was changed by several hundreds of nanometers and the topography and the electronic properties of the graphene sample were studied. By using *in situ* prepared tips, one can repeatedly obtain atomic resolution images without observing changes in the quality of the images (Figure 3a). Different regions of the graphene/Ru (0001) unit cell were selected to perform tunneling spectroscopy (see the marks in Figure 3b). Figure 3c-e shows the whole set of 256 differential conductance *vs.* voltage (d$I$/d$V$ *vs. V*) traces, obtained by numerical differentiation of current *vs.* voltage traces, in a two-dimensional histogram [29] which shows the average d$I$/d$V$ *vs. V* trace and its dispersion at the three selected locations of the Moiré pattern. The average d$I$/d$V$ *vs. V* trace is also shown in the 2D histogram. The low dispersion of datapoints around the averaged trace proves the high reproducibility obtained with the *in situ* prepared tips which is a consequence of the structural stability of the tip mechanically annealed tip apex.

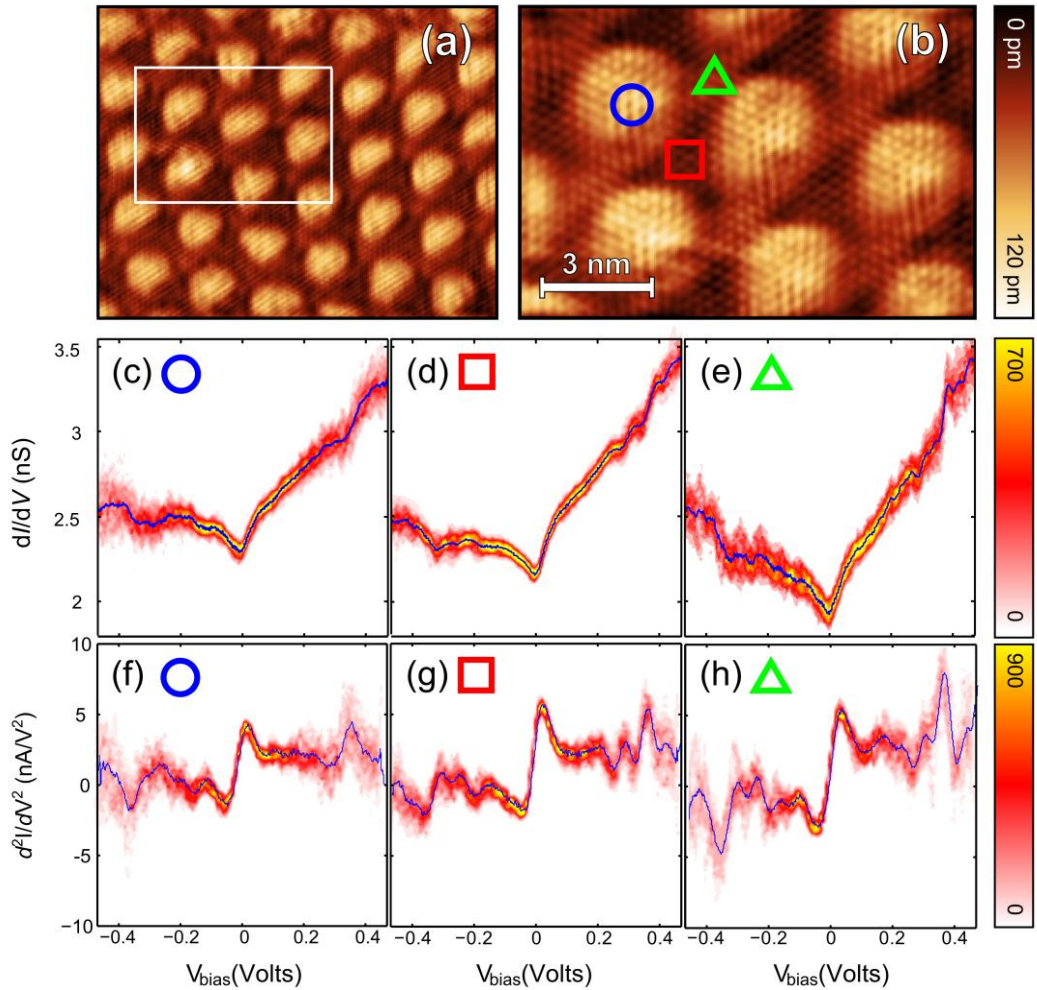

**Figure 3**: (a) STM topography of a large area epitaxial graphene on Ru (0001) forming a Moiré pattern. The atomic corrugation is also visible superimposed to the Moiré pattern ($V_{sample}$ = 1 V; $I$ = 1.5 nA). (b) A high resolution STM topography of the region marked by the rectangle in (a) ($V_{sample}$ = 1 V; $I$ = 1.5 nA). Atomic corrugation can be more easily resolved here. (c)-(e) two dimensional histograms of the differential conductance as a function of the tip bias voltage built from the numerical differentiation of 256 $I(V)$ traces measured at the regions marked by the circle, square and triangle in (b). To build these 2D histograms both the bias voltage and the d$I$/d$V$ axes are discretized into $N$ bins forming an $N$ by $N$ matrix (256 by 256 in this case). Each data point whose d$I$/d$V$ and $V$ values are within the interval of one bin, adds one count to it. The number of counts in each bin is then denoted with a color scale. The averaged d$I$/d$V$ trace is also plotted on top of the 2D histogram (solid blue line). (f)-(g) same as (c)-(e) but showing the second derivative of the current d$^2I$/d$V^2$. The peaks at ± 370 mV indicate an inelastic tunneling process such as the excitation of phonons in the graphene lattice by electron-phonon interaction.

The d$I$/d$V$ spectra are reproducible and different at the hills and the valleys of the rippled graphene. The conductance at the Fermi level $E_F$ is 20% larger on top of the hills. Additionally, in both valleys and hills there is a prominent increase of conductance (up to a 5-10%) around a bias voltage of ± 370 meV. This type of feature, commonly observed in inelastic tunneling spectroscopy (IETS), is associated to the excitation of phonons in the lattice. This feature can be more easily resolved by plotting the second derivative of the current (d$I^2$/d$V^2$ *vs. V*), appearing as a positive (negative) peak at positive (negative) voltage (see Figure 3f-h).

In order to further test the robustness of the tips prepared *in situ*, tunneling spectroscopy measurements have been carried out with different tips prepared following the same procedure described in previous sections. Figure 4 shows the tunneling spectra measured at the valley of the rippled graphene using three different tips. The two-dimensional histograms are built from 256 differential conductance *vs.* voltage (d$I$/d$V$ *vs.* $V$) traces. The spectroscopic data measured with different tips are in remarkable agreement and show low dispersion around the average d$I$/d$V$ *vs.* $V$. Indeed the three studied valleys shows a very similar asymmetry in their shape and present a well-defined minimum conductance value at zero bias voltage and very similar increase of differential conductance around ± 370 meV bias voltage. The small variations from trace to trace are attributed to the large dependence of the tunneling spectra on the exact position of the tip on the sample. This demonstrates the reproducibility and stability of the tips prepared *in situ* with the procedure presented here.

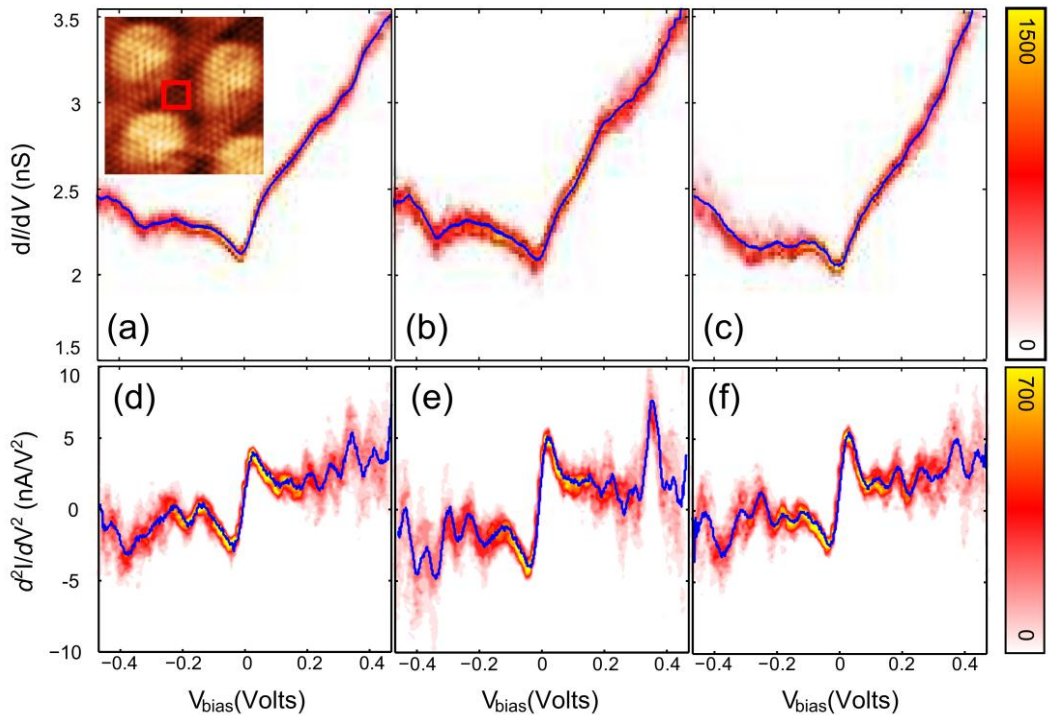

**Figure 4**: (a)-(c) Two dimensional histograms of the differential conductance as a function of the tip bias voltage built from the numerical differentiation of 256 $I(V)$ traces measured with three different *in situ* prepared tips at a valley of the Moiré pattern (see inset). Notice the reproducibility of the measured spectra despite they correspond to different tips, which share the same *in situ* preparation procedure. (d)-(f) same as (a)-(c) but showing the second derivative of the current d$^2I$/d$V^2$.

## 5. Conclusions

In summary, we have presented an *in situ* tip preparation procedure compatible with scanning tunneling microscopes operated at cryogenic temperatures and high magnetic fields. This procedure does not require additional UHV setups such as annealing or ion milling stages. Our procedure relies on the electric-field-induced deposition of material from the tip. A mechanical annealing treatment, by performing repeated indentations onto the sputtered material, is used to form a highly crystalline tip apex structure which increases the stability of the scanning tunneling microscopy and spectroscopy measurements. We employed *in situ* prepared tips to study the topography and the electronic properties of epitaxial graphene grown on Ru (0001), at low temperature < 300 mK, demonstrating that the use of these *in situ* prepared tips increases the stability of the scanning tunneling images and the reproducibility of the spectroscopic measurements.

**Acknowledgements**

This work was supported by Comunidad de Madrid (Spain) through the program NANOBIOMAGNET (CAM s2009/MAT-1726), MICINN (Spain) through the programs MAT2008-01735, MAT2011-25046 and CONSOLIDER-INGENIO-2010 CSD-2007-00010 and by the EU through the network "ELFOS" (FP7-ICT2009-6).

# Supplementary material:

# Highly reproducible low temperature scanning tunnelling microscopy and spectroscopy with *in situ* prepared tips

Andres Castellanos-Gomez[1,2]*  Gabino Rubio-Bollinger[1,3,]*, Manuela Garnica[1,4], Sara Barja[1,4], Amadeo L. Vázquez de Parga[1,3,4], Rodolfo Miranda[1,3,4] and Nicolás Agraït[1,3,4].

**Conductance histogram measured on a deposited nanodot:**

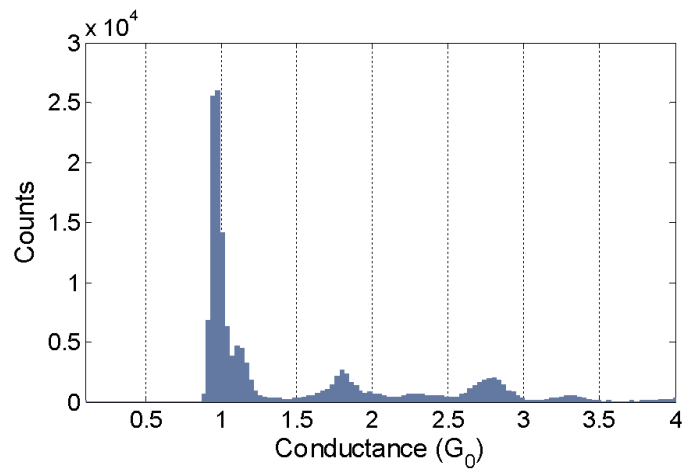

**Figure S1**: Conductance histogram of gold contacts performed on a gold nanodot, sputtered from the tip by field-effect-induced deposition, during the mechanically annealing of the tip prepared *in situ*. The histogram has been built from 400 consecutive conductance *vs.* distance traces.

[1] Departamento de Física de la Materia Condensada (C–III). Universidad Autónoma de Madrid, Campus de Cantoblanco, 28049 Madrid, Spain.
[2] Kavli Institute of Nanoscience, Delft University of Technology, Lorentzweg 1, 2628 CJ Delft, The Netherlands.
[3] Instituto Universitario de Ciencia de Materiales "Nicolás Cabrera". Campus de Cantoblanco, 28049 Madrid, Spain.
[4] Instituto Madrileño de Estudios Avanzados en Nanociencia IMDEA-Nanociencia, 28049 Madrid, Spain.
*E-mail: a.castellanosgomez@tudelft.nl , gabino.rubio@uam.es